\documentclass[reprint, amsmath, amssymb, prb]{revtex4-1}

\usepackage{graphicx}  
\usepackage{dcolumn}   
\usepackage{bm} 
\usepackage{amssymb}   
\graphicspath{{Pictures/}} 
\usepackage{booktabs}
\usepackage{graphicx}
\usepackage{epsfig,amsmath,amssymb,color,float,setspace}
\usepackage[utf8]{inputenc}
\usepackage[T1]{fontenc}
\usepackage{titlesec}
\usepackage{booktabs}
\usepackage{booktabs}
\usepackage[bottom]{footmisc}
\usepackage{multirow}

\usepackage{hhline}

\definecolor{myblue}{RGB}{0, 112, 192}

\begin{document}

\preprint{APS/123-QED}

\title{Prediction of a Heusler alloy with switchable metal-to-half-metal behavior}


\author{Vasiliy~D.~Buchelnikov$^{1,2}$}
\author{Vladimir~V.~Sokolovskiy$^{1,2}$}
\author{Olga~N.~Miroshkina$^{1,3,4}$}
\author{Danil~R.~Baigutlin$^{1,3}$}
\author{Mikhail~A.~Zagrebin$^{1,2,5}$}
\author{Bernardo~Barbiellini$^{3,6}$}
\author{Erkki~L\"{a}hderanta$^{3}$}
\affiliation{$^{1}$Faculty of Physics, Chelyabinsk State University, 454001 Chelyabinsk, Russia}
\affiliation{$^{2}$National University of Science and Technology "MISiS", 119049 Moscow, Russia}
\affiliation{$^{3}$Department of Physics, School of Engineering Science, LUT University, FI-53850 Lappeenranta, Finland}
\affiliation{$^{4}$Faculty of Physics and Center for Nanointegration Duisburg-Essen (CENIDE), University of Duisburg-Essen, 47048 Duisburg, Germany}
\affiliation{$^{5}$National Research South Ural State University, 454080 Chelyabinsk, Russia}
\affiliation{$^{6}$Physics Department, Northeastern University, Boston, Massachusetts 02115, USA}

\date{\today}

\begin{abstract}
We propose a ferromagnetic Heusler alloy that can switch between a metal and a half-metal. This effect can provide tunable spintronics properties. Using the density functional theory~(DFT) with reliable implementations of the electron correlation effects, we find Mn$_2$ScSi total energy curves consisting of distinct branches with a very small energy difference.
The phase at low lattice crystal volume is a low magnetic half-metallic state while the phase at high lattice crystal volume is a high magnetic metallic state. 
We~suggest that the transition between half-metallic and metallic states 
can be triggered by a triaxial contraction/expansion of the crystal lattice or by an external magnetic field if we assume that the lattice is cubic and remains cubic under expansion/contraction. However, the phase at high volume can also undergo an austenite-martensite phase transition because of the presence of Jahn-Teller active $3d$ electrons on the Mn atoms.

\end{abstract}

\maketitle

\section{INTRODUCTION}
Nowadays, spintronics is a rapid-developing field of science and technology~\cite{Gallagher2019,Fert2008,dieny2020,puebla2020}, which aims to use the intrinsic spin of the electron and its associated magnetic moment in solid-state devices~\cite{Bhatti-2017,Hirohata2020} including synchronized networks of spin-transfer oscillators~\cite{Torrejon2017}, spin-transfer torque~\cite{Liao2020} and magnetoelectric random access memory devices~\cite{Zhang2019,Wu-2011}, spin transfer nano-generators~\cite{Fu-2012}, and spin holographic processors~\cite{Fetisovs-2018}.
The~efficiency of these devices is related directly to the level of spin injection from the electrodes to the semiconductors and to their degree of spin polarization~\cite{Ram-2020}.
Half-metallic~(HM) ferromagnetic~(FM) compounds are characterized by an energy gap in one spin direction at the Fermi level. 
They therefore exhibit metallic character in one spin-channel and semiconducting behavior in the other spin-channel~\cite{Bainsla2016, Shakil-2020}.

Among HM ferromagnets, the half- and full-Heusler alloys are of great interest because they usually demonstrate stable half-metallicity with high Curie temperature and  spin polarization~\cite{Palmstrom-2003, Mori-2012, Bainsla2016}. 
Nowadays, much attention is given to the wide variety of FM Co$_2YZ$ ($Y =$~Fe, Mn and $Z =$~Si, Ge, Sn)~\cite{Hoat2018, Deka-2014, Siakeng2018, Umetsu2012, Comtesse20015, Zagrebin_2016, Meinert_2012_prb, Miroshkina-2020,Galanakis-2006effect}, Fe$_2YZ$ ($Y=$~Cr, Mn, Co and $Z =$~Si, Al, Ga)~\cite{Chaudhuri2019, He2017, Hongzhi-2007,Ram-2020}
and ferrimagnetic~(FIM) Mn$_2YZ$ ($Y=$~V, Cr, Fe, Co, Ni and $Z=$~Al, Ga, Si, Ge, Sn, In)~\cite{Luo-2008_JMMM, Zenasni-2013, Luo-2008_JAP, Sokolovskiy-2015, Ram-2020, Chadov-2010, Graf-2010, Nayak-2015} that were studied both theoretically and experimentally.  
In~the Co$_2YZ$ family, the HM behavior appears when the $Y$ element has less valence electrons than~Co. 
On~the other hand, the half-metallicity in Fe$_2YZ$ and Mn$_2YZ$ families  occurs regardless the number of valence electrons in the $Y$ atoms. 
Therefore, the HM behavior is promoted by the Mn and Fe atoms. 
Moreover, parallel or antiparallel spin alignment of two Mn atoms can lead to FM and FIM order and to noncolinear configurations.
Since the Mn$^{3+}$ ion has a $d^4$ electronic configuration with  partially filled $e_g$ orbitals, it is Jahn-Teller active.
This effect explains the tetragonal distortion producing high magnetocrystalline anisotropy~(MCA). 
This observation makes FIM Mn$_2$-based alloys useful for transfer torque application, where high MCA is a key factor for fast switching with low currents and high thermal stability~\cite{Felser-2015, winterlik2012}.


An~interesting idea to add Sc (with one $d$ valence electron) into the matrix of the  Mn$_2$-based family has been recently proposed by Ram~\textit{et~al.}~\cite{Ram-2020}.
These authors performed  DFT calculations for  Mn$_2$Sc$Z$ ($Z=$~Si, Ge, Sn)  Heusler alloys taking in account on-site Coulomb interaction effects.
The results indicated HM behavior with a narrow band gap in the case of Mn$_2$ScSi and Mn$_2$ScGe whereas Mn$_2$ScSn displayed metallic behavior.
Ram~\textit{et~al.}~\cite{Ram-2020} claimed that one must use an accurate DFT scheme to reproduce their results since
the deficiencies of traditional DFT exchange correlation 
approximations such as the local spin density approximation (LSDA)
and the generalized gradient approximation
(GGA) prevent the appearance of HM properties in Mn$_2$Sc$Z$ alloys.
Therefore, one must reduce the self interaction errors  
present in LSDA and GGA  by using schemes such as the DFT$+U$ method \cite{Cococcioni2005}. 
Recently, we have studied correlation effects beyond the GGA in $\alpha$-Mn~\cite{Pulkkinen-2020},  Ni$_2$- and  Co$_2$-based Heusler alloys~\cite{Buchelnikov-2019,Miroshkina-2020} by using the  strongly constrained and appropriately normed~(SCAN) functional~\cite{Sun-2015} (meta-GGA method), which contains  self-interaction correction without introducing an
explicit Hubbard parameter~$U$.

In~this work, we confirm that reliable corrections beyond the GGA  not only stabilize the HM properties of the Mn$_2$ScSi alloys in a robust way but reveal a surprisingly rich phase diagram useful for tunable spintronics applications. 
For example, one could produce spintronic logic devices with switches that can operate on femtosecond time scales \cite{tengdin2020}.
Our results show that Mn$_2$ScSi is an exemplar magnetic functional material hosting competing phases. The lattice of this material provides an elastic environment where charge, spin, and orbital degrees of freedom interact and produce unexpected functionalities~\cite{Kakeshita2011}.
Small energy differences between phases also offer unique 
opportunities to benchmark Coulomb correlation effects in~DFT.

The outline of the paper is as follows. Section~II contains the calculation details. Section~III is devoted to the discussion of the results of structural, magnetic, and electronic properties and phase stability of Mn$_2$ScSi.
The concluding remarks are presented in Sec.~IV.

\section{CALCULATION DETAILS}
To~perform the calculations, we employed the PAW method implemented in VASP code~\cite{Kresse-1996,Kresse-1999} using 16-atom supercells.
The~GGA for the exchange correlation functional was treated within the Perdew, Burke, and Ernzerhof~(PBE)~\cite{Perdew1996} scheme. 
Electron correlation effects beyond GGA were included using both GGA$+U$ by Dudarev~\textit{et~al.}~\cite{Dudarev_1998} and meta-GGA SCAN by Sun~\textit{et~al.}~\cite{Sun-2015}. 
The~parameter $U$ for Mn is taken in the interval from 0.2 to 2~eV. 
The~value $U_{\mathrm{Mn}}=1.973$~eV was chosen in accordance with  Ram~\textit{et~al.}~\cite{Ram-2020}. 
Since the Coulomb correlation for  Sc affects weakly  the electronic and magnetic properties, we choose $U_{\mathrm{Sc}}=0.435$~eV proposed by Ram~\textit{et~al.}~\cite{Ram-2020}. 
The~volume optimization of regular and inverse Heusler structures (space groups $Fm\overline{3}m$ and $F\overline{4}3m$) with different magnetic order is perform.
For~all functionals, the geometry optimization procedure yields the regular L2$_1$-cubic Heusler structure as the most favorable one with the FIM order involving antiparallel alignment of Mn and Sc magnetic moments (see Supplementary Material~(SM)~\cite{SM}).

\section{RESULTS AND DISCUSSIONS}

\begin{figure}[!t]
\includegraphics[width=\columnwidth]{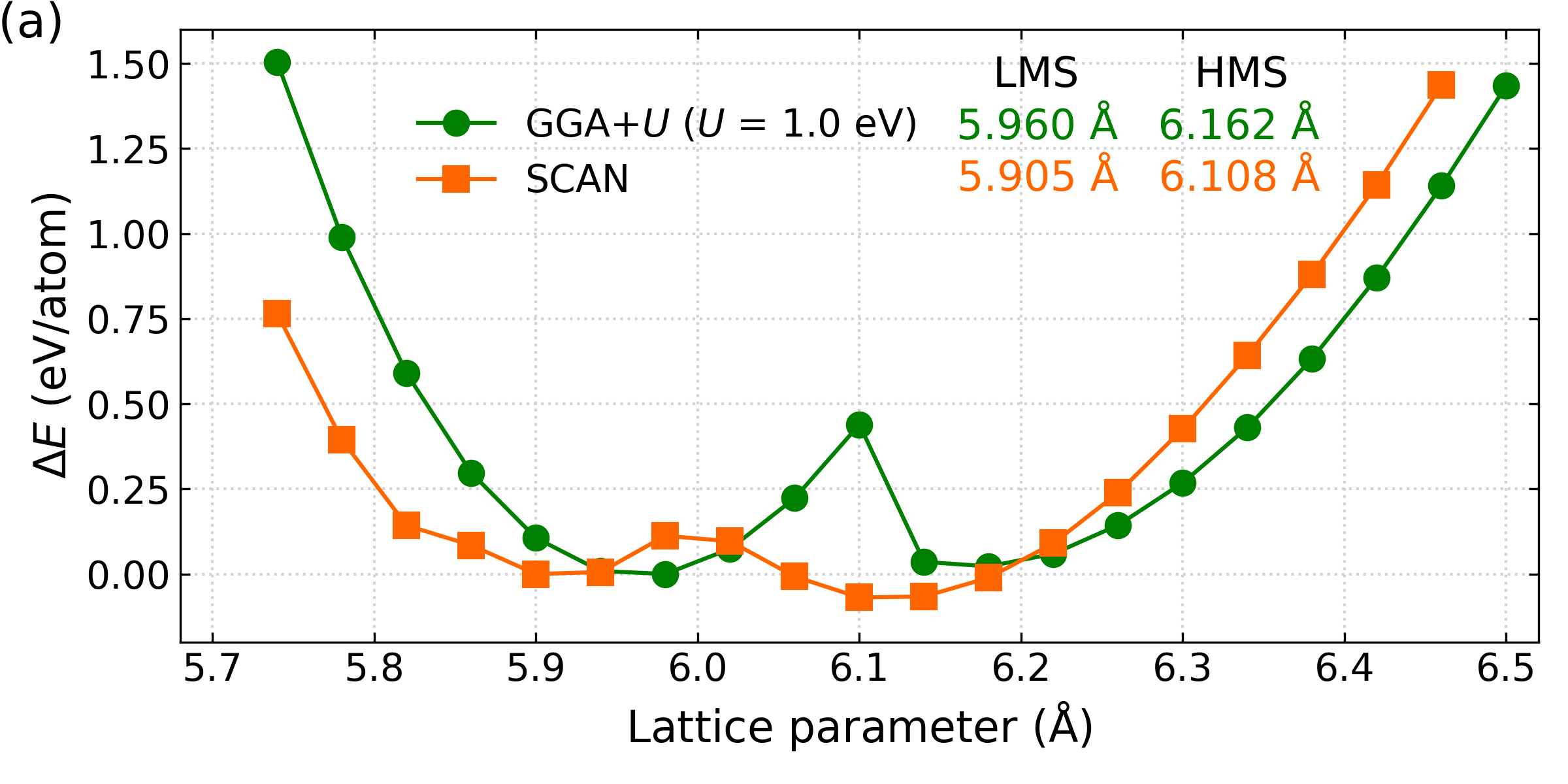} 
\includegraphics[width=\columnwidth]{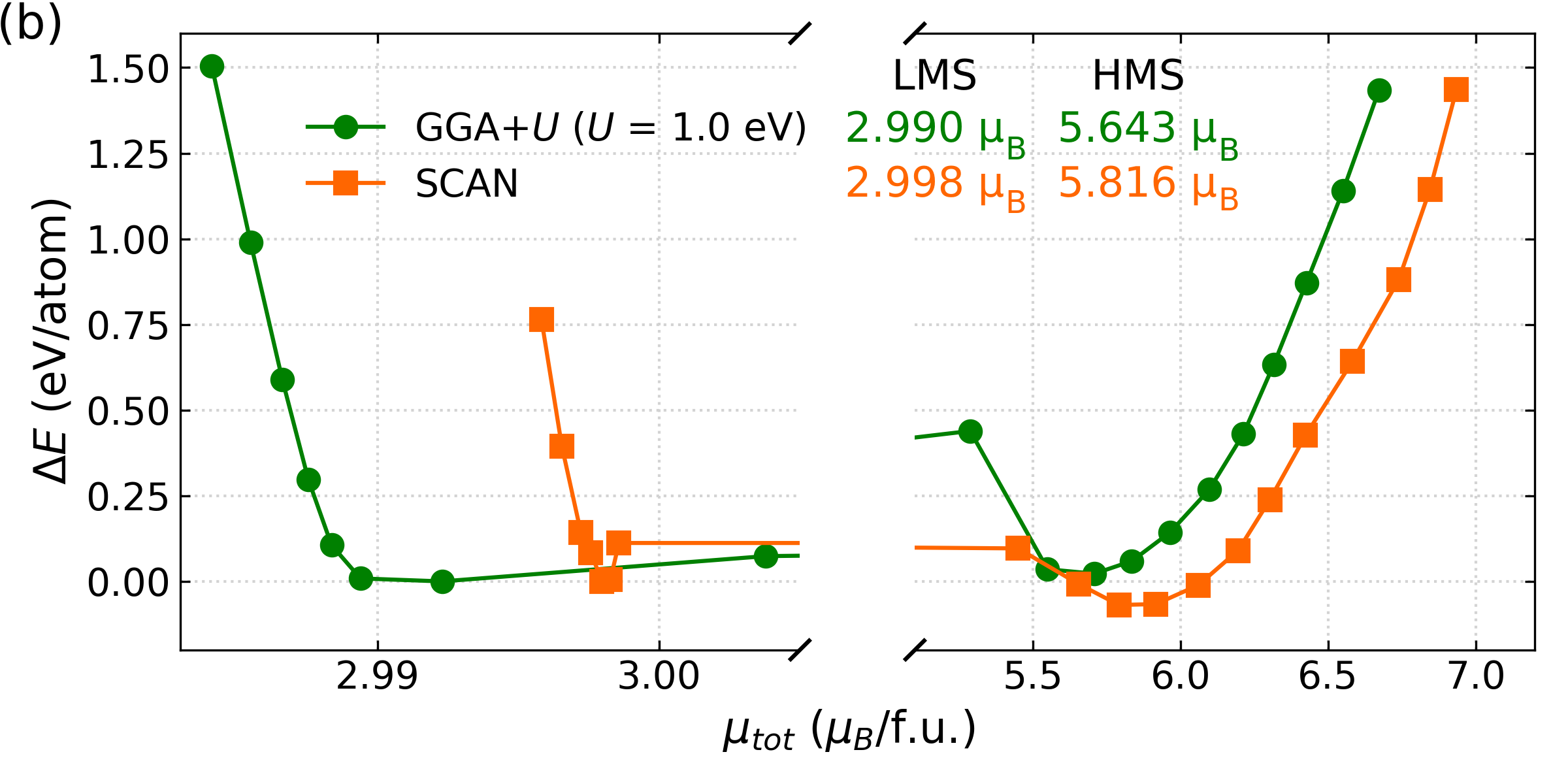}
\caption{ The total energy difference~($\Delta E$) as a function of (a)~lattice parameter and (b)~magnetic moment of Mn$_2$ScSi for SCAN and GGA$+U$ ($U = 1$~eV) solutions.
 For~each cases, the $\Delta E$ is plotted with respect to the left energy minimum.  }
\label{Fig-1}
\end{figure}
\begin{figure}[!t]
\includegraphics[scale=0.58]{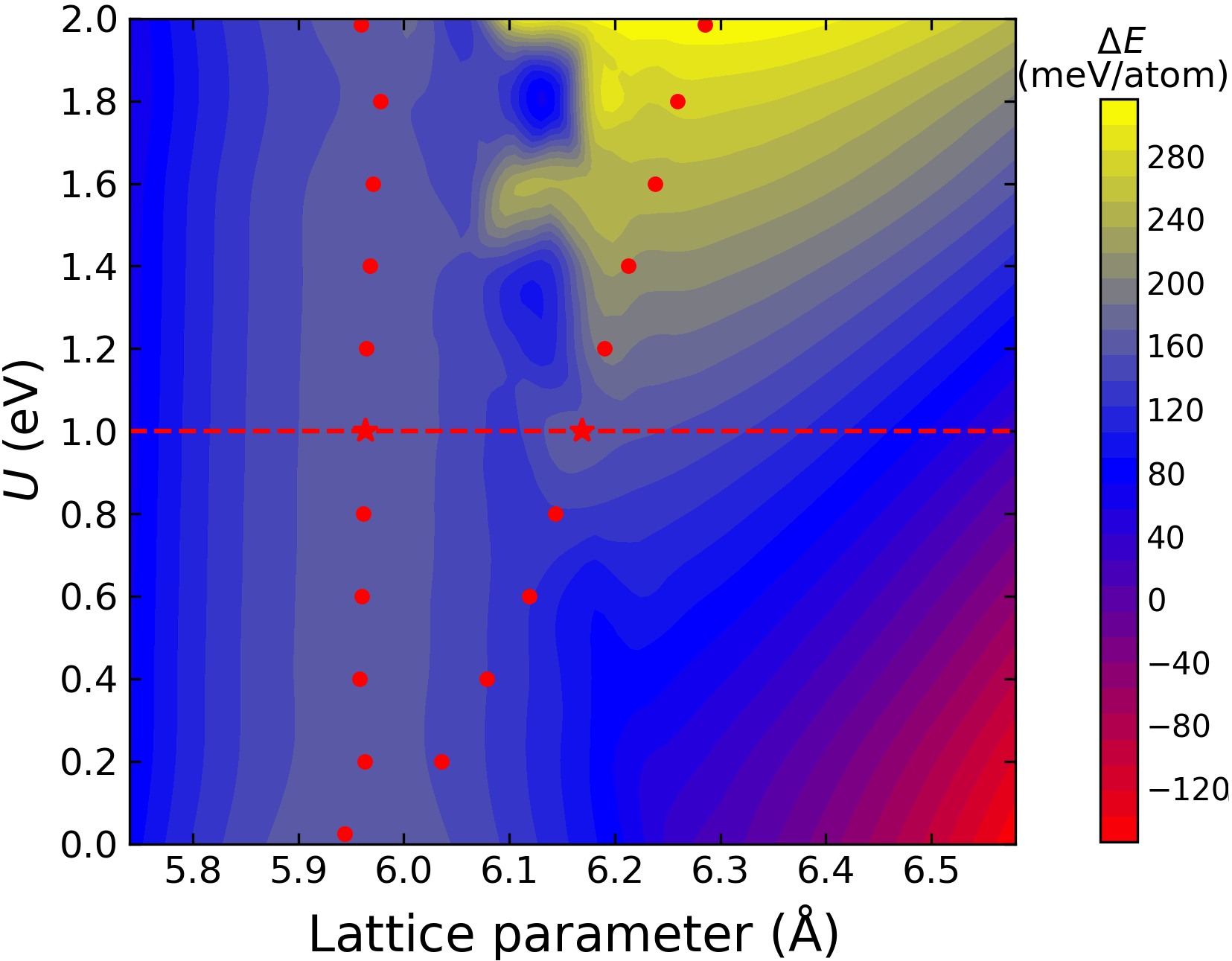}
\caption{ The~total energy difference for Mn$_2$ScSi
calculated by GGA$+U$ for a set of $U$ values and mapped into the diagram <<Coulomb repulsion term~($U$)~-- lattice parameter~($a$)>>. 
$\Delta E$~is plotted with respect to the minimum for~LMS. 
The~optimized lattice parameters, which are estimated from the fitting for the Birch-Murnaghan equation of state for both LMS and HMS, are marked by the red symbols.
 The~stars denote degeneracy of the ground state for which the LMS and HMS have a similar energy at~$U=1$~eV.
}
\label{Fig-2}
\end{figure}
\begin{figure*}[!t]
\includegraphics[scale=0.4]{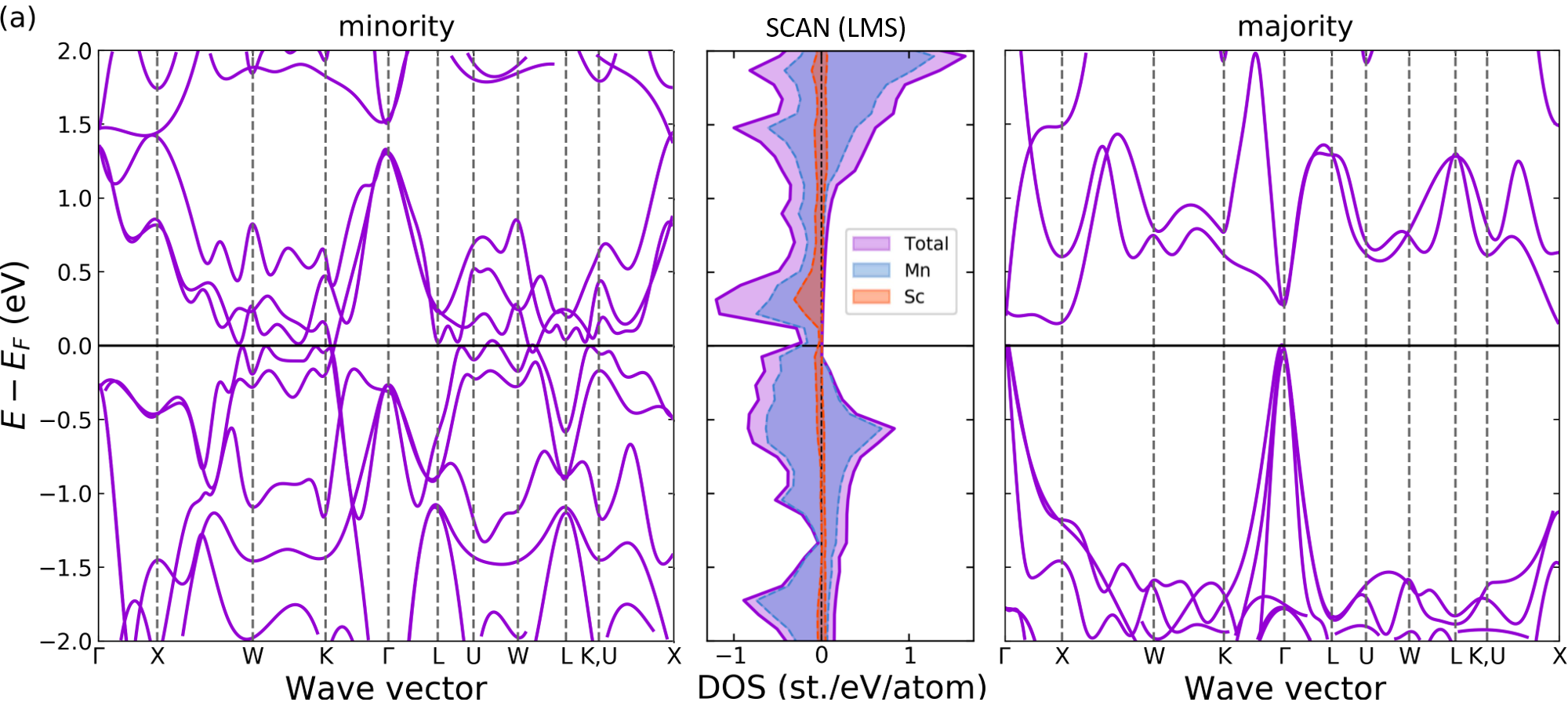}
\\
\includegraphics[scale=0.4]{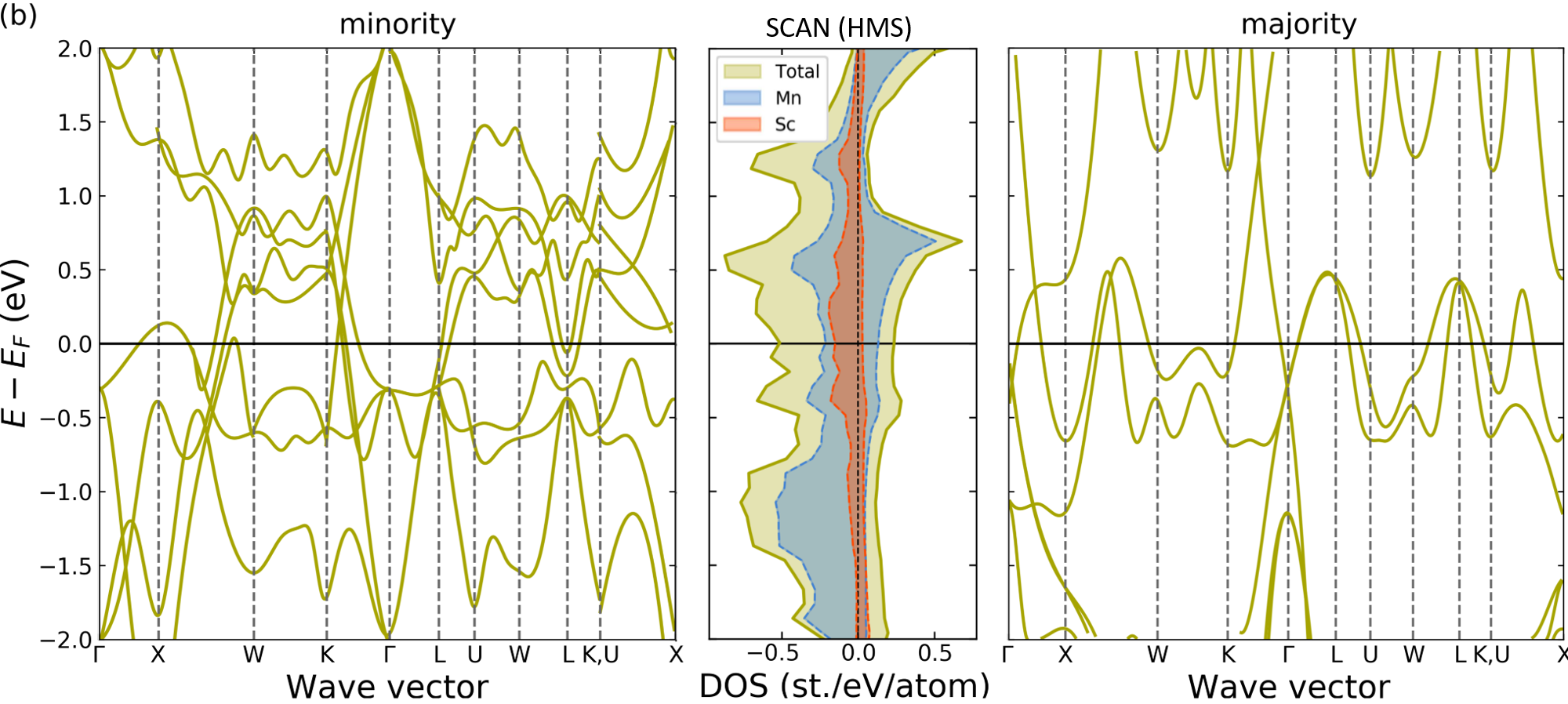}
\caption{SCAN energy band structure, total and partial DOSs of Mn$_2$ScSi calculated for the (a)~LMS and  (b)~HMS with the optimized lattice constants $a_0^{\mathrm{LMS}} = 5.905$~\AA~ and $a_0^{\mathrm{HMS}} = 6.108$~\AA, respectively. }
\label{Fig-3}
\end{figure*}

\subsection{Structural optimization}
We~first consider the total energy results. 
In~Fig.~\ref{Fig-1}, we  show the total energy as a function of the  lattice constant~($a$) and the total magnetic moment~($\mu_{tot}$) calculated within SCAN and GGA$+U$. 
SCAN yields a degenerated ground state with two almost equal energy minima observed at lattice constants of 5.905 and 6.108~\AA. 
These two minima correspond to two magnetic states with the values of $\mu_{tot}$ equal to 3 and 5.8~$\mu_B$/f.u. as shown in Fig.~\ref{Fig-1}(b).
We~denote these states as the low-magnetic state~(LMS) and high-magnetic state~(HMS), respectively.
The~energy and magnetic moment differences between LMS and HMS are $\Delta E = 3.75$~meV/atom and $\Delta \mu = 2.90~\mu_B$/f.u., respectively. 
The total magnetic moment in SCAN is an integer in agreement with the empirical Slater-Pauling~(SP)~\cite{SP} rule thereby revealing the HM behavior.



In contrast to SCAN, our GGA calculations give only one clear minima at lattice constant~$a_0=5.94$~\AA.~
Ram~\textit{et~al.}~\cite{Ram-2020} have only considered GGA$+U$ corrections around the GGA total energy minima but they have missed the second total energy minima at higher volume. 
In our case, the volume optimization with 
fixed $U$ value of 1.973~eV as Ram~\textit{et~al.}~\cite{Ram-2020} yields 
both a local energy minimum at a similar equilibrium volume as GGA
and the global minimum at larger volume as illustrated in the SM~\cite{SM}.

To~understand better these results, we performed a  parametric study
of the Hubbard parameter $U$. Figure~\ref{Fig-2} illustrates a contour map of the total energy as functions of $U$ and lattice parameter $a$. The set of $E(a)$ curves for various $U$ is given in the~SM~\cite{SM}. We delineate in this contour map shown in Fig.~\ref{Fig-2} a triangle linking optimized lattice constants at local and global energy minima corresponding to the LMS and HMS, respectively. The vertex of this triangle is located close to $U$ equal zero demonstrating that GGA is a singular point in terms of correlation effects. As soon as we introduce a small $U$ value, two well defined structures with different magnetization appear in the energy landscape. 
Related parametric studies in $\gamma$-Mn by Podloucky and Redinger~\cite{Podloucky-2018} and by Pulkkinen~\textit{et~al.}~\cite{Pulkkinen-2020} have found that $U$ approximately 1~eV gives the corrected equilibrium volume. Therefore, we believe that ~$U \approx 1$~eV is a correct energy scale to describe the correlation effects in the present case. 
For~$U = 1$~eV, the LMS and HMS are found to be close to each other in the energy  as shown in Fig~\ref{Fig-1}(a) in agreement with SCAN.

The transition between LMS and HMS can be achieved by a uniform contraction/expansion of the crystal lattice by  $\approx 3.3$~\%. 
We~predict that the magnetovolume effect~\cite{magnetovolume} ($\frac{\Delta V}{V_{\mathrm{HMS}}}$) should be accompanied by in the change in magnetization ($\Delta \mu_{tot}$) corresponding to $\approx$2.8~$\mu_{B}$/f.u. as it is derived from SCAN calculations. For GGA+$U$ with $U = 1$~eV, $\Delta \mu_{tot}$ is slightly less and equals to $\approx$2.7~$\mu_{B}$/f.u.
Another way is to switch the transition by applying magnetic field.
\subsection{Electronic structure}
To understand the electronic structure,
we consider the spin-polarized band structures and the density of states~(DOS). The bands are calculated along the high-symmetric points of the first Brillouin zone for the majority and minority spin channels
for the GGA, GGA$+U$, and SCAN methods. 
In~Fig.~\ref{Fig-3}, we present both the SCAN bands and DOS curves calculated at the optimized lattice constants for LMS and HMS found in Fig.~\ref{Fig-1}(a). 
The corresponding figures for GGA and GGA$+U$  are given in the SM~\cite{SM}.  As~Fig.~\ref{Fig-3}(a) suggests, in the case of LMS, the minority-spin bands present few band crossings at the Fermi level ($E_F$), while the majority-spin bands reveal an clear energy gap around $E_F$. 
Such HM behavior for Mn$_2$ScSi is quite different of that in regular HM Heusler alloys in which the spin up states are filled and the spin down states are unoccupied at $E_F$~\cite{Felser-2015}.
 The energy gap at $\Gamma$ point for the majority bands is direct as shown in Fig.~\ref{Fig-3}(a) and its~predicted value within SCAN is 0.32~eV which is twice the value calculated by Ram~\textit{et~al.}~\cite{Ram-2020}. 
 

In~terms of the HMS (Fig.~\ref{Fig-3}(b)), both the minority and majority spin bands as well as DOSs show clear metallic behavior. 
It is important to notice that the bands calculated for the GGA$+U$ with $U = 1.973$~eV at the optimized volume~\cite{SM} display metallic character in contrast to the bands reported by Ram~\textit{et~al.}~\cite{Ram-2020} indicating that the HM properties is lost for the larger volume. Therefore, it is possible to switch the HM behavior by a uniform contraction/expansion of the crystal lattice.
 
\begin{figure}[hbt]
\includegraphics[scale=0.4]{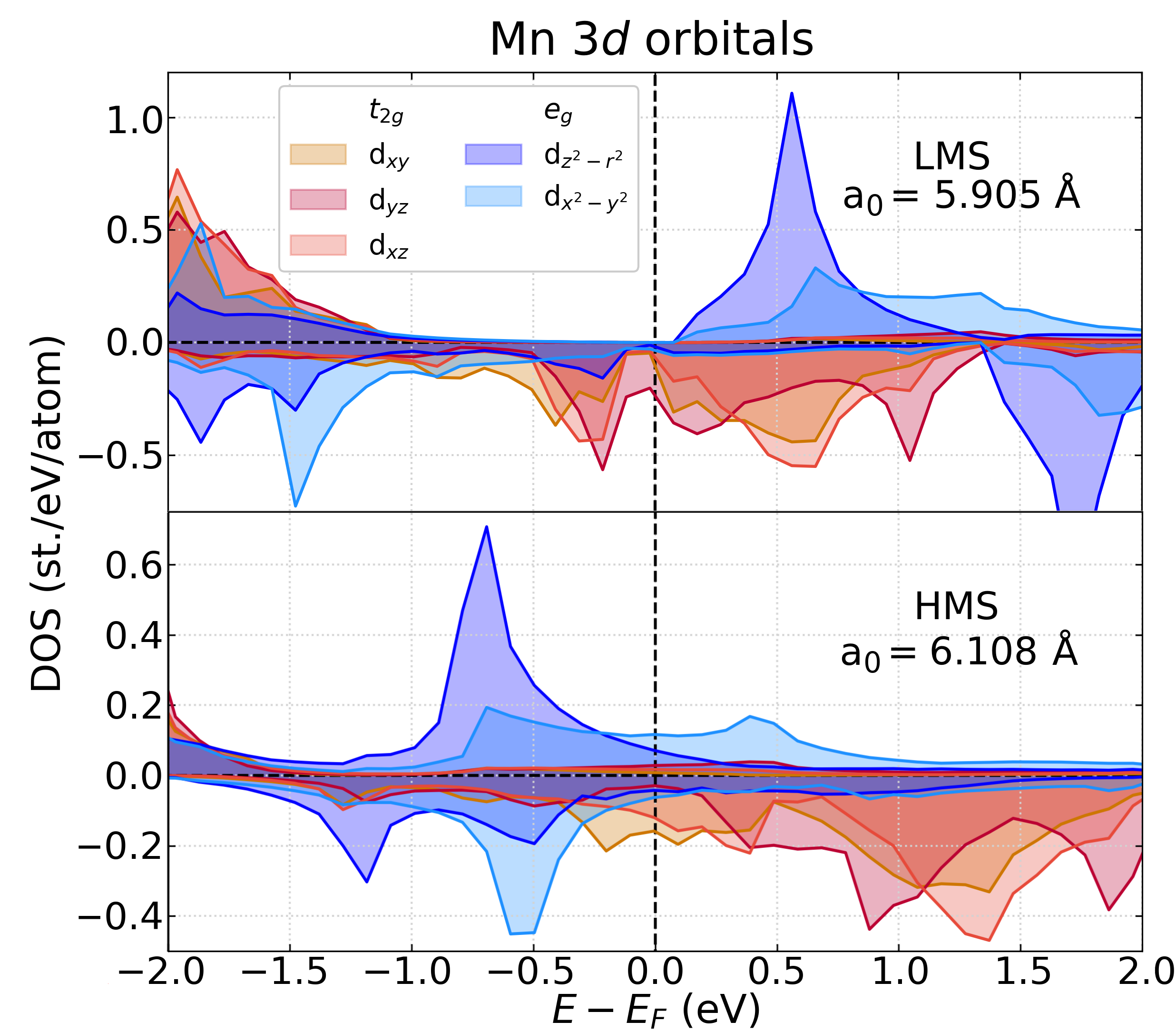} 
\caption{ Mn-3$d$ orbital resolved DOSs for LMS (upper) and HMS (bottom) calculated with SCAN.}
\label{Fig-4}
\end{figure}

Such switching behavior can be rationalized by analyzing the 3$d$ partial DOSs for Mn atoms in the vicinity of $E_F$ shown in Fig.~\ref{Fig-4}. For~the LMS, three $t_{2g}$ orbitals at $E_F$ are present in the minority spin channel and contributed to the integer magnetic moment, whereas two $e_g$ orbitals are almost empty. In contrast, both the $t_{2g}$ and $e_g$ orbitals at $E_F$ for the minority spin channel are occupied for the~HMS.
For the majority spin channel we observe mostly $e_g$ orbitals. 
The main difference between LMS and HMS is the occupation of an energy $e_g$ peak for the majority spin band. When the $e_g$ state is occupied the system undergoes an austenic-martensite transition because of Jahn-Teller effects.

Our results above apply if the lattice is cubic and remains cubic under triaxial  expansion/contraction~\cite{Relax}.
To~examine the martensitic phase, we have performed total energy calculations as a function of tetragonal distortion ratio $c/a$ assuming a constant volume between austenite ($c/a = 1$) and martensite ($c/a \ne 1$). 
Indeed the martensitic phase with $c/a = 1.27$ is found to be energetically favorable~\cite{SM}. 
The predicted  $\Delta \mu_{tot}$ between austenitic and martensitic phase is about 1.807~$\mu_{B}$/f.u. This finding allows us to conclude that Mn$_2$ScSi should also display magnetocaloric properties in the vicinity of magnetostructural phase transformation \cite{Planes2009}. 

\subsection{Phase stability}
\begin{figure}[!t]
\includegraphics[scale=0.43]{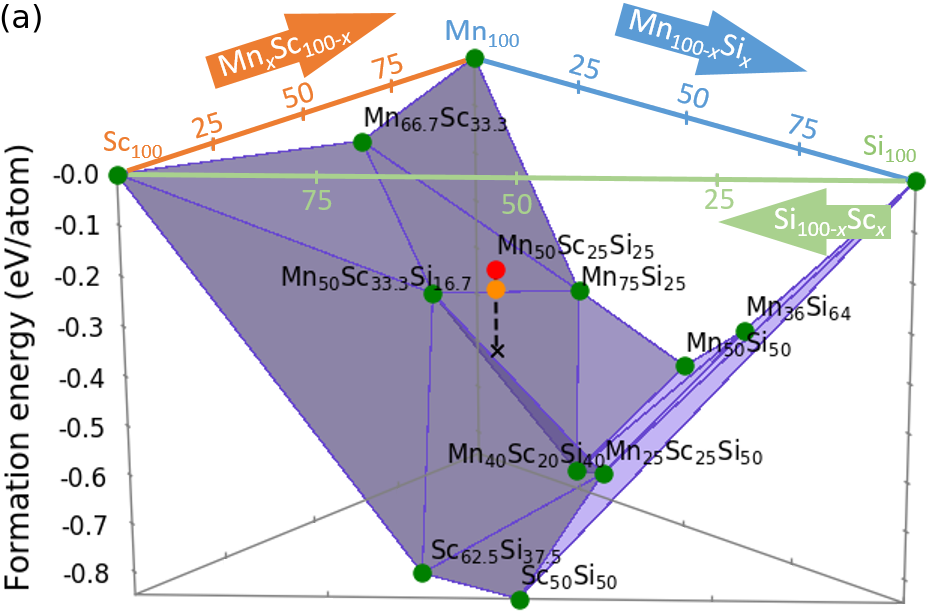}
\includegraphics[scale=0.52]{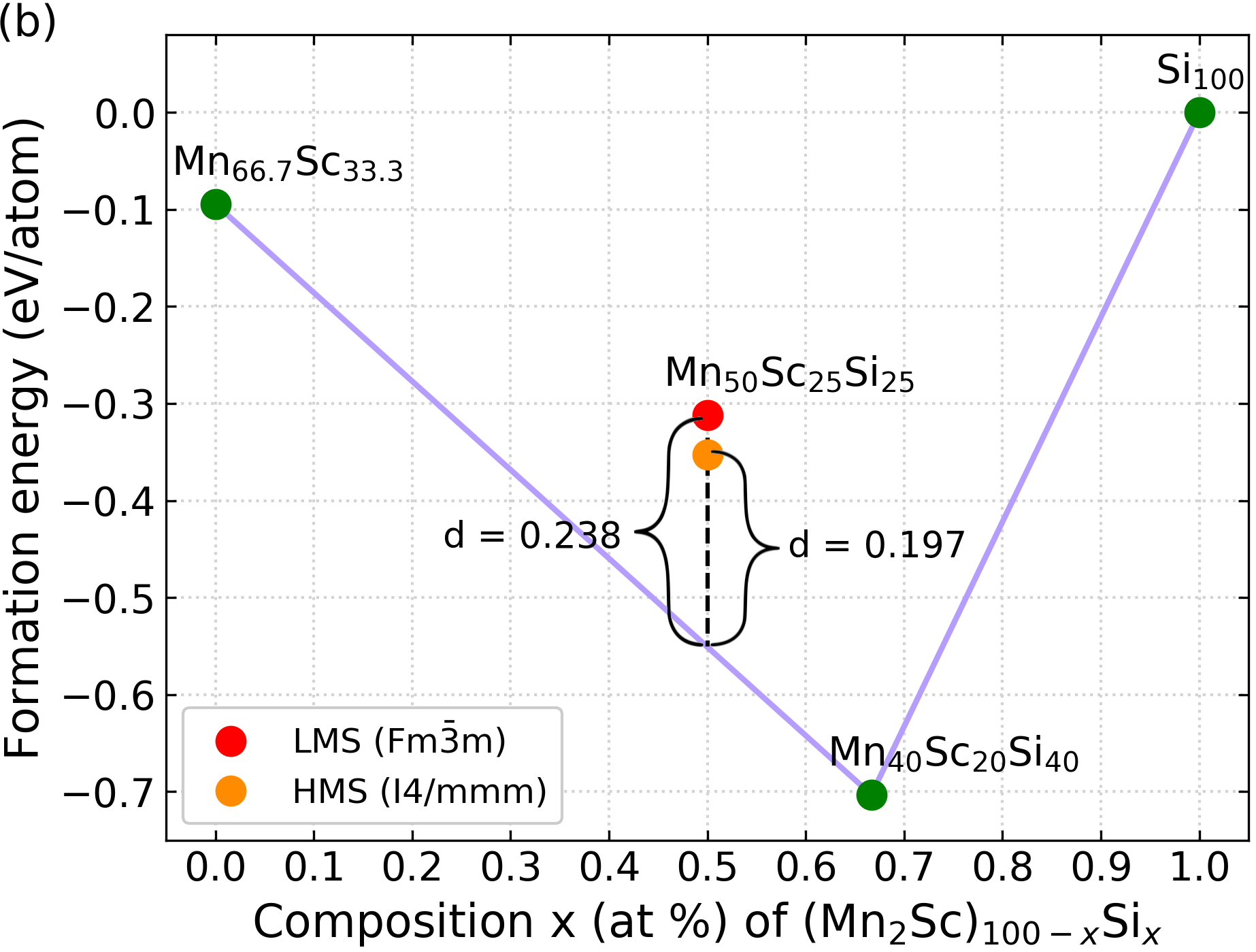}
\caption{ (a)~The hull energy convex and (b)~cross section of hull energy convex that contains the convex hull distance for Mn$_2$ScSi  with the LMS cubic structure and HMS tetragonal one. The~formation energies are calculated with SCAN.}
\label{Fig-5}
\end{figure}

Since the predicted Mn$_2$ScSi compound 
has not yet been grown, it is important to evaluate its phase stability.
We have therefore checked the thermodynamic stability in three steps.

As a preliminary test, we have performed formation energy ($E_{form}$) calculations for Mn$_2$ScSi with respect to the elemental components in their groundstate bulk structures. The negative values of $E_{form}^{\mathrm{SCAN}}$ ($-0.312$ and $-0.319$~eV/atom for LMS and HMS, respectively) indicate thermodynamic stability of Mn$_2$ScSi in the corresponding cubic states~\cite{SM}.

However, stability with respect to the sum of total energies of the corresponding pure elements is a necessary but not yet a sufficient
condition for thermodynamic phase stability. In order to clarify this 
point, we must compare also $E_{form}$ of a compound against to all stable combination of phases at that composition. This aim can be achieved 
with the convex hull construction for the phase space of interest~\cite{Oses2019aflow}. Typically, a convex hull connects stable phases that are lower in $ E_ {form} $ than any other phase under consideration in this overall composition. In this way, phases located above the convex hull are metastable or unstable while phases placed on the convex hull are stable.
The pivot points ( 12 stable phases) for the convex hull of the Mn-Sc-Si ternary system
were taken according to the AFLOW database~\cite{Oses2019aflow} (they are listed in Table~2 of SM~\cite{SM}). By considering the $E_{form}^{\mathrm{SCAN}}$ of these phases, we constructed the three-dimensional convex hull shown in Fig.~\ref{Fig-5}(a). Here,~we
indicate $E_{form}^{\mathrm{SCAN}}$ for Mn$_2$ScSi with the LMS cubic structure and HMS tetragonal one. The formation energies 
of pure elements are set to zero. 
We show the cross section of the hull energy convex in Fig.~\ref{Fig-5}(b)
to estimate the distance from the most stable phase. 
The Mn$_2$ScSi is located above the convex hull by $\approx$0.24~eV/atom for LMS cubic phase as shown in Fig.~\ref{Fig-5}(b).
Interestingly, this distance is about 0.188~eV/atom~\cite{Oses2019aflow}~(http://www.aflowlib.org)
for GGA-PBE. Thereby, this result suggests a 
metastable tendency of the
Mn$_2$ScSi compound. 

In order to understand this metastable behavior, 
we compare the Mn$_2$ScSi phase energy to the energies of decomposition products (pivot points of the convex hull) at that composition and calculate the mixing energy $E_{mix}$. 
To~accomplish this task, we consider 23 possible decomposition reactions into the three stable components (see Table~3 and Fig.~7 in SM~\cite{SM}). As~a result, we found that 9 of 23~reactions yield a negative sign $E_{mix}$ indicating the stability of LMS cubic phase against to segregation process. Therefore, we conclude that Mn$_2$ScSi 
can be grown as a metastable compound. 
We must keep in mind that the well-known Ni$_{2}$Mn$_{1+x}Z_{1-x}$ Heusler 
alloys show both
experimentally and theoretically a segregation tendency into ternary stoichiometric and binary compounds~\cite{Yuhasz2010, Ccakir2017, Krenke2016, Sokolovskiy2019} due to several-step heat treatment. 
Despite this metastable character, Ni$_2$Mn-based alloys still exhibit remarkable multi-functional properties.

\section{CONCLUSION}
In~conclusion, accurate DFT calculations in Mn$_2$ScSi Heusler alloy reveals an almost equal double energy minima behavior for close lattice constants but different magnetic moments. Mn$_2$ScSi displays either half-metallic (in LMS) or metallic (in HMS) behavior as a function of the lattice parameter. 
Our~study shows how one can preserve the HM phase or one can switch to the metallic~ phase.
We suggest that the half-metallic $\leftrightarrow$ metallic transition might be 
realized by applying an external magnetic field or pressure. The critical magnetic field evaluated from the relation between $\Delta E$  and Zeeman energy  is about 11~T. 
While
the magnetovolume effect and critical pressure are predicted to be  3.3\% and 20 GPa, respectively. 
A second interesting aspect of the correlation effects is related to the prediction of an austenite-martensite transformation for the metallic phase, which produces a change in magnetization and a magnetocaloric effect.

We suggest that the switching mechanism between  half-metallic and metallic behavior could be implemented for the design  of spintronics devices like spin filters, sensors, switches, logical gates with femtosecond time scale~\cite{tengdin2020}. Till date many studies have been performed theoretically and experimentally to investigate the externally controlled carrier’s spin polarization and half-metallic--metallic transition in HM Heusler alloys by applying pressure, magnetic or electric fields~\cite{Zayed2018, Zhou2015, Zhang2015, Miroshkina-2020}. However, in all these cases, the external load was applied for a certain length of time, which implies significant energy consumption. 
 In the present work, the proposed switching behavior in 
Mn$_2$ScSi can be realized at once.
Since this external perturbation is limited in time,
the energy consumption is significantly lowered.
This finding opens new avenues for fast and energy efficient spintronics applications.

\section{ACKNOWLEDGMENTS}
SCAN calculations were supported by the RSF~-~Russian Science Foundation project No.~17-72-20022. GGA calculations were performed with the support of the Ministry of Science and Higher Education of the Russian Federation within the framework of the Russian State Assignment under contract No.~075-00250-20-03. GGA+$U$ calculations were funded by the RFBR~-~Russian Foundation for Basic Research No.~20-42-740006.
B.B.~acknowledges CSC-IT Center for Science, Finland, for computational resources and support from the COST Action~CA16218.
O.N.M.~acknowledges the Deutsche Forschungsgemeinschaft (DFG, German Research Foundation)~- Project-ID 405553726~- TRR~270, subproject~B06. V.D.B. and M.A.Z. gratefully acknowledge the ﬁnancial support of the Ministry of Science and Higher Education of the Russian Federation in the framework of Increase Competitiveness Program of NUST “MISiS” (No. K2-2020-018), implemented by a governmental decree, N 211.

\label{References}

\bibliography{literature}

\begin{thebibliography}{62}
\bibitem [{SP()}]{SP}%
  \BibitemOpen
\bibfield  {journal} {  }\href@noop {} {\bibinfo  {journal} {According to the
  SP rule for HM systems, the total magnetic moment is equal to $\mu_{tot} =
  N_e - 24$, where $N_e$ is the total number of valence electrons. For
  Mn$_2$ScSi, $N_e$ is 21, and so $\mu_{tot} = 3 \mu_B$/f.u. by absolute value
  is expected. Notice, the negative sign of $\mu_{tot}$ yields the energy gap
  at the Fermi level for majority density of states}\ }\BibitemShut {NoStop}%
\bibitem [{Rel()}]{Relax}%
  \BibitemOpen
\bibfield  {journal} {  }\href@noop {} {\bibinfo  {journal} {We have checked
  the stability of LMS and HMS in a cubic phase to verify that the solution
  converges to the cubic lattice by considering a full structural relaxation
  using supercells with artificially broken symmetries. Our calculations have
  shown that the cubic symmetry is preserved for both LMS and~HMS if the
  relaxation calculations are performed in the vicinity of corresponding
  optimized lattice constants~$a_0$}\ }\BibitemShut {NoStop}%
\end{thebibliography}
\end{document}